\documentclass[intlimits,twoside,a4paper]{article}

\usepackage{amsmath,amssymb}
\usepackage{graphicx}
\usepackage{wrapfig}

\usepackage[T2A]{fontenc}
\usepackage[cp1251]{inputenc}

\usepackage{cmpj2}

\issue{2013}{16}{1}{13001}

\doinumber{10.5488/CMP.16.13001}

\title{ Density of one-particle states for 2D electron gas in magnetic field}

\author[I.M. Dubrovskyi]{I.M. Dubrovskyi\thanks{E-mail: dubrovskii@voliacable.com}}

\authorcopyright{I.M. Dubrovskyi, 2013}

\address{Institute for Metal Physics,
36 Vernadsky St., Kyiv 03680, Ukraine}

\date{Received July 3, 2012, in final form  September 25, 2012}

\begin{document}

\maketitle
\begin{abstract}
The density of states of a particle in a 2D area is independent
both of the energy and  form of the area only at the region of large
values of energy. If energy is small, the density of states in the
rectangular potential well
essentially depends on the form of the area. If the bottom of the potential well has a potential
relief, it can define the small eigenvalues as the discrete
levels. In this case, dimensions and form of the area would not
have any importance. If the conservation of zero value of the
angular momentum is taken into account, the effective one-particle
Hamiltonian for the 2D electron gas in the magnetic field in the
circle is the Hamiltonian with the parabolic potential and the
reflecting bounds. It is supposed that in the square, the
Hamiltonian has the same view. The 2D density of states in the
square can be computed as the convolution of 1D densities. The
density of one-particle states for 2D electron gas in the
magnetic field is obtained. It consists of three regions. There is
a discrete spectrum at the smallest energy. In the intervening
region the density of states is the sum of the piecewise
continuous function and the density of the discrete spectrum. At
great energies, the density of states is a continuous function.
The Fermi energy dependence on the magnetic field is obtained when
the field is small and the Fermi energy is located in the region
of continuous spectrum. The Fermi energy has the  oscillating correction and in the
average it increases proportionally to the square of the magnetic
induction. Total energy of electron gas in magnetic field also
oscillates and increases when the
magnetic field increases  monotonously.


\pacs 05.30.Ch, 75.20.-g
\end{abstract}

\section{Introduction}
The spectrum of one-particle Hamiltonian in a bounded area is
discrete. When values of energy are more than maximum of potential
energy (in the fourth section this criterion will be improved),
the distances between levels are of the order
$\hbar^{2}/2mW^{2/D}$. Here, $m$ is the mass of the particle, $W$
is the volume of the area, and $D$ is the dimensionality of space.
When the volume of the area is macroscopic, the spectrum may be
considered as quasicontinuous.Then, the density of states
$\mathfrak{N}_{D}(E)$ may be introduced:
\begin{equation}\label{1}
\mathfrak{N}_{D}(E)=\frac{N_{D}(E+\delta E)-N_{D}(E)}{\delta E}\,.
\end{equation}
Here, $N_D(E)$ is the number of states, energy eigenvalues of which
are less than $E$, $\delta E$ is a small interval of energy which
is larger than the distances between discrete levels. Function $N_D(E)$
was considered mathematically rigorously in the monograph
\cite{1}. The Schr\"{o}dinger equation was multiplied by
$2m/\hbar^{2}$, and the eigenvalue $\varepsilon = 2mE/\hbar^{2}$
has the dimension of physical quantity $[L^{-2}]$ (for brevity let
us also refer to it as ``energy''). It was shown that the asymptotical
formulae at $\varepsilon \to \infty$ for functions
$N_{D}(\varepsilon)$ are:
\begin{equation}\label{2}
N_{2}(\varepsilon)<\frac{S}{4\pi}\varepsilon +\theta c \sqrt
{\varepsilon},  \qquad
N_{3}(\varepsilon)<\frac{V}{6\pi^{2}}\varepsilon^{3/2}+\theta c
\varepsilon,
\end{equation}
when the wave functions are equal to zero at the boundaries of the
area. Here, $\vert \theta \vert <1$ and $c$ is independent of
$\varepsilon$. It is important that the first terms of these
formulae are independent of the form of the area. The functions
$N_{D}(\varepsilon)$ take only integer values. If this is neglected
when energy is great, the formula (\ref{1}) can be considered as
$\mathfrak{N}_{D}=\partial N_{D}/\partial \varepsilon$. Then, we
can obtain from the formulae (\ref{2}):
\begin{equation}\label{3}
\mathfrak{N}_{2}=\frac{S}{4\pi}\,,\qquad
\mathfrak{N}_{3}=\frac{V}{4\pi^2}\sqrt{\varepsilon},
\end{equation}
if only the first terms of the asymptotical expansions are taken
into account. Evidently, extrapolation of the formulae (\ref{3}) to
small values of energy is incorrect even when there is no potential energy.

In the second section of this work, the density of states at small
energy is considered in the absence of a potential energy.

In the third section of this work the 2D density of states is
considered in the potential well that is created by the harmonic
potential and the reflecting boundaries. As is shown in the work
\cite{2}, this potential takes place in the effective one-particle
Hamiltonian of the electron gas in the magnetic field. The
dependencies on the magnetic field are investigated for the Fermi
energy and for the total energy of the gas.

The fourth section is a mathematical supplement. The problem on
the linear harmonic oscillator with condition zeroes of wave
function at the segment ends is considered.

\section{The density of states in a square}
The derivation of the formula (\ref{2}) for $N_{2}(\varepsilon)$
starts from consideration of the square with $S=L^2$ (see
monograph \cite{1}). The states are determined by two integer
numbers. The energy of state $|kl\rangle$ is
$\varepsilon_{kl}=\pi^2(k^2+l^2)/L^2$. Then, $N_{2}(\varepsilon)$
will be equal to the number of junctions of the net of squares that are
parallel to the coordinate axes and have the sides equal to unit,
which fit into the interior of the positive quadrant of the circle
with radius $(L/\pi)\sqrt{\varepsilon}$. This quantity differs
from the area of the quadrant $S\varepsilon /4\pi$ by the sum of
areas of partial squares that are crossed by the circle. The second
term in the formula (\ref{2}) is the approximate estimate of this
amendment. The relative magnitude of this amendment will be
smaller, when the quadrant radius is larger.

Let us consider the other method of calculating the state density
in the quadratic area that can be used for small energy values
too. The 2D Schr\"{o}dinger  equation for a free particle, provided
 that the wave function is equal zero at the boundaries of the
square, can be changed by two identical 1D equations. An
eigenvalue of the 2D equation is the sum of eigenvalues of 1D
equations. Therefore, let us consider the state density for the 1D
equation.

The number of states, whose eigenvalues are less than
$\varepsilon$ for 1D equation, is as follows: $N_{1}(\varepsilon)=\lceil L
\sqrt{\varepsilon}/\pi\rceil$. Here, $\lceil \alpha \rceil$ denotes
the integer part of the number $\alpha$. The density of states that
is determined by formula (\ref{1}) is an interval function rather
than a point function. The magnitude of the interval cannot be
taken arbitrarily small. In 1D, this magnitude is limited by the demand
that one eigenvalue should be in the interval at the greatest energy
$\varepsilon_{\mathrm{m}}$. Then,
\begin{equation}\label{4}
\frac{L}{\pi}\left(\sqrt{\varepsilon_{\mathrm{m}}+\delta}-\sqrt{\varepsilon_{\mathrm{m}}}\right)\geqslant
1,\qquad \delta \geqslant   \left(\frac{\pi}{L}\right)^2
+\frac{2\pi}{L}\sqrt{\varepsilon_{\mathrm{m}}}\approx
\frac{2\pi}{L}\sqrt{\varepsilon_{\mathrm{m}}}\,.
\end{equation}
In most cases, the density of states is used in integral formulae.
Then, an interval function can be changed by a piecewise continuous
stepped function or a continuous differentiable function that is
determined by any method of interpolation. It is apparent that the
consideration of peculiarities of the state density is
meaningless.

Let us determine the state density for the square at the values of
energy that are smaller than $\varepsilon_{\mathrm{m}}$. The interval
$\delta$ is determined by the formula (\ref{4}), and it is
accepted as the unity of energy. Non-dimensional ($\varepsilon
/\delta $) eigenvalues of energy for 1D equation are denoted as
$\lambda_{1}$ and $\lambda_{2}$, and non-dimensional eigenvalue of
energy for 2D equation is denoted as $\mu ,\; \mu = \lambda_{1} +
\lambda_{2}$. Let us consider the intervals
$[\lambda_{1},\lambda_{1}+1]$ and $[\lambda_{2}-1,\lambda_{2}]$
where $\lambda_{2}=\mu - \lambda_{1}$. The eigenfunctions of the
2D equation that are products\; of\; the\; eigenfunctions 1D
equations, which are related to these intervals, have the
eigenvalues that are located at interval $[\mu - 1,\mu + 1]$. The
number of these states is denoted as $M_{2}(\mu - 1,\mu + 1)$. In a
similar way, the eigenvalues of the 2D equation that are located
at the interval $[\mu ,\mu + 2]$ are obtained when
$\lambda_{2}=\mu - \lambda_{1} + 1$. Then, the number of the
eigenvalues of the 2D equation at the interval $[\mu,\mu + 1]$
(when $\mu \geqslant  2$) is:
\begin{eqnarray}\label{5}
M_{2}(\mu,\mu + 1)&=& \frac{1}{2}\sum_{\lambda = 0}^{\mu -
1}M_{1}(\lambda ,\lambda + 1)M_{1}(\mu -\lambda -1,\mu -\lambda )
\notag\\&&{}+\frac{1}{2}\sum_{\lambda = 0}^{\mu }M_{1}(\lambda ,\lambda +
1)M_{1}(\mu -\lambda ,\mu -\lambda +1).
\end{eqnarray}
It follows from the formula (\ref{4}) that:
\begin{equation}\label{6}
M_{1}(\lambda ,\lambda + 1)=\frac{L\sqrt{\delta}}{\pi} \left(
\sqrt{\lambda +1} - \sqrt{\lambda} \right) .
\end{equation}
Then, $M_{2}(\mu,\mu + 1)=(L^2 \delta /2\pi ^2)\varSigma (\mu )$,
where
\begin{eqnarray} \label{7}
\varSigma (\mu )&=&-\sum_{i=1}^{\mu -1}\left(\sqrt{i} \right)\left(\sqrt{\mu -i+1}
-\sqrt{\mu -i-1}\right)\notag\\&&{}+ \sum_{i=0}^{\mu -1}\left(\sqrt{i+1} \right)\left(\sqrt{\mu
-i+1} -\sqrt{\mu -i-1}\right)+\sqrt{\mu +1}-\sqrt{\mu }.
\end{eqnarray}
This function of integer argument can be written as:
\begin{equation} \label{8}
\varSigma (\mu )=S(\mu +1)-S(\mu )-S(\mu -1)+S(\mu -2),
\end{equation}
where
\begin{equation} \label{9}
S(x)= \sum_{n=0}^{x-1} \sqrt{(n+1)(x-n)},\qquad x\geqslant  1;\qquad
S(0)=0.
\end{equation}
The function $S(x)$ can be computed using Euler-Maclaurin method:
\begin{equation} \label{10}
S(x)= \left(\frac{x+1}{2}\right)^2
\arcsin{\frac{x-1}{x+1}}+x\sqrt{x} -\frac{x-1}{12\sqrt{x}} \,.
\end{equation}
Based on the approximate formula
\begin{equation} \label{11}
\arcsin{\frac{x-1}{x+1}}=\frac{\pi}{2}-\frac{2}{\sqrt{x+1}}-\frac{1}{3(x+1)^{3/2}}+\frac{9}{4(x+1)^{5/2}}\,,
\end{equation}
the asymptotical formula
\begin{equation} \label{12}
\varSigma (\mu )\approx \frac{\pi}{2}+\frac{3}{4\sqrt{\mu}}
\end{equation}
can be obtained. The function $\varSigma (\mu )$ can be calculated numerically. The results are obtained from the
formula (\ref{7}) and from the asymptotical formula (\ref{12})
tabulated in table~1.
The values of the function $\varSigma (\mu )$ approach $\pi
/2=1.57096$ from above.
\begin{table}[h]
\caption{The function $\Sigma(\mu)$  that is numerically calculated [formula (7)], and calculated using an asymptotical formula (12).}
\vspace{2ex}
\begin{center}
 \begin{tabular}{|l|l|l|l|l|l|l|l|}\hline\hline
 $\mu$       & 1    & 10   & 100   & 1000  & 10,000  & 100,000 & 1,000,000 \\ \hline\hline
formula (7)& 1.81 & 1.577&1.571 &1.5708&1.570797&1.570796&\\
\hline
 formula (12)& & &1.65 & 1.59 &1.578 &1.573 &1.5715\\ \hline\hline
\end{tabular}
\end{center}
\end{table}

For sufficiently large value $\mu$ the density of states in a
square is described by formula:
\begin{equation} \label{13}
\mathfrak{N} _{2}(\varepsilon =\mu \delta )=\frac{M_{2}(\mu ,\mu
+1)}{\delta}=\frac{L^2}{2\pi ^2}\varSigma (\mu )\approx
\frac{L^2}{4\pi}+\frac{3L^2\sqrt{\delta}}{8\pi ^2
\sqrt{\varepsilon}}\,.
\end{equation}
The first term in this formula coincides with
the common expression for 2D system. It can be obtained
by differentiation $N_{2}(\varepsilon )$ [formula (\ref{2})],
where $S=L^2$. This formula can be used for derivation of the
density of states for a flat geometrical figure of arbitrary shape
(see monograph~\cite{1}). In this process, $L^2$ is changed by the
figure area $S$ and amendments are proportional to
$\varepsilon^{-1/2}$, i.e., they alter the second term in the
formula (\ref{13}). Therefore, the second term depends on the
figure shape and on the interval magnitude~$\delta$.

The sign of the second term also depends on the figure shape. In
the square this term is positive, i.e., when the energy increases,
the density of states decreases tending to the constant value from
above. This is explained by the fact that in 1D, the state density
increases when the energy decreases. The eigenfunctions in a
circle are the Bessel functions of the first kind
$J_{n}(r\sqrt{\varepsilon})$. The eigenvalues in this case are
$\varepsilon_{nk} =j_{nk}^2/R^2$ where $R$ is the circle radius,
and $j_{nk}$ is the null of the function $J_{n}$ that has the
number $k$ in the order of increasing. There is no formula that
describes these nulls when their numbers $k$ are small, but it is
known that the distances between nulls increase when their numbers
decrease. Therefore, the density of states should decrease when the
energy decreases, and the amendment should be negative.

In fact the spectrum at small energy values is formed by the
potential relief of the bottom of the potential well. The
distances between energy levels are determined by parameters of
this relief. Therefore, this spectrum cannot be considered as
quasicontinuous. The density of states in this case can be
described by the set of $\delta$-functions. By virtue of the fact
that the determining factor is the potential relief, it is believed
that the figure shape is of no significance. Then, it may be helpful to obtain the state density for the square in the whole region of energy values.

If the 2D Schr\"{o}dinger equation with the potential energy can
be solved by separating the variables in the Cartesian
coordinates, then the 2D density of states can be obtained. Every
interval $[\varepsilon_{1} , \varepsilon_{1}+ \rd\varepsilon]$ on
the axis of energy of 1D states contains
$\mathfrak{N}_{1}(\varepsilon_{1})\rd\varepsilon$ states, whose wave
functions are $\psi (x_{1})$. Products of these functions
with the wave functions $\psi (x_{2})$ that have the energy values
in the interval $[\varepsilon_{2} , \varepsilon_{2}+
\rd\varepsilon]$, where $\varepsilon_{2} =\varepsilon
-\varepsilon_{1}$, are the wave functions of the 2D states, the
energies of which are in the interval $[\varepsilon,\varepsilon
+\rd\varepsilon ]$. The number of these states connected with
the energy value $\varepsilon_{1}$ is:
\begin{equation} \label{14}
\rd M_{2} (\varepsilon,
\varepsilon_{1})=\mathfrak{N}_{1}(\varepsilon_{1})
\mathfrak{N}_{1}(\varepsilon -\varepsilon_{1})\rd\varepsilon
\rd\varepsilon .
\end{equation}
Then, the density of states in the square is:
\begin{equation} \label{15}
\mathfrak{N}_{2}(\varepsilon )=\frac{\rd M_{2}(\varepsilon
)}{\rd\varepsilon}=\int_{0}^{\varepsilon}\mathfrak{N}_{1}(\alpha)
\mathfrak{N}_{1}(\varepsilon -\alpha)\rd\alpha ,
\end{equation}
i.e., it is a convolution of 1D densities of states.

The form of the spectrum in the region of small energy values can
play a significant role  depending on the form of the
potential relief of quantities that are determined by integral
formulae.

\section{The density of states and energy of 2D electron gas in the magnetic field}
It is shown in the work \cite{2} that the statistical operator of
the electron gas in the magnetic field is defined by effective
Hamiltonian that is the sum of the same one-particle Hamiltonians.
Each one-particle Hamiltonian describes the particle in the
potential well with a harmonic potential and reflecting boundaries.
Electrons interact with each other and with the neutralizing
background. The electron density in the magnetic field should be
distributed in such a way as to shield the harmonic potential. It
is shown in the work \cite{2} that this shielding in the circle
with\; radius\; $R$ \;leads \;to \;renormalization\; of the
electron charge $e_{\mathrm{r}}\backsim e\sqrt{a_{0}/R}$, where $(-e)$ is
the electron charge, $a_{0}$ is the Bohr radius. The residual
harmonic potential is proportional to $\omega^2$ where $\omega =
e_{\mathrm{r}}H/m$ is the cyclotron frequency, $H$ is the magnetic
induction, $m$ is the electron mass.

Let us suppose that in the square with the side $2L$ and zero of
coordinate system in the center, the effective one-particle
Hamiltonian with the symmetrical gauge also has the view:
\begin{equation} \label{16}
\hat{h}=-\frac{\hbar^2}{2m}\left( \frac{\partial^2}{\partial
x^2}+\frac{\partial^2}{\partial
y^2}\right)+\frac{m\omega^2}{8}\left(x^2+y^2\right).
\end{equation}
Here, $ e_{\mathrm{r}} =e\sqrt{a_{0}/L}$. Separating the variables and
multiplying by $2m/\hbar^2$, we obtain two identical 1D
equations:
\begin{equation}\label{17}
\psi''+\frac{2mE_{\nu}}{\hbar^2}\psi-\frac{m^2\omega^2x_{\nu}^2}{4\hbar^2}\psi=0,\qquad
\nu=1,2.
\end{equation}
The boundary conditions are:
\begin{equation}\label{18}
\psi (\pm L)=0.
\end{equation}
The problem on a linear oscillator with the boundary condition
(\ref{18}) is considered in the fourth section. In this case, the
1D density of states is as follows:
\begin{eqnarray}\label{19}
\mathfrak{N}_{1}(\varepsilon
)&=&\sum_{n=0}^{n_{0}}\delta(\varepsilon-\varepsilon_{n})+\Theta(\varepsilon-\varepsilon_{\mathrm{b}})\frac{L}{\pi
\sqrt{\varepsilon}}\,,\notag\\
\varepsilon_{n}&=&\Delta\left(n+\frac{1}{2}\right),\qquad
\varepsilon_{\mathrm{b}}=\Delta\left(n_{0}+\frac{1}{2}\right),\qquad
n_{0}\approx\left \lceil \frac{4m\omega
L^2}{\pi^2\hbar}\right\rceil ,\qquad \Delta
=\frac{m\omega}{\hbar}\,.
\end{eqnarray}
Here, the first term is the spectrum of the linear oscillator
without reflecting boundaries. (We change $n+\gamma(n)$ to $n$
because $\gamma(n)$ are practically everywhere small). For
$\varepsilon>\varepsilon_{\mathrm{b}}$, the density of states is the density
of quasicontinuous spectrum. Boundary magnitude $n=n_{0}$ is
determined approximately, but it is significant that $n_{0}$ is
proportional to the magnetic induction and it is an integer.

The density of states of Hamiltonian (\ref{16}) can be obtained by
the formula~(\ref{15}). It consists of three regions. At the
smallest energy, the spectrum is discrete:
\begin{subequations}
\begin{equation}
\mathfrak{N}_{2\mathrm{d}}(\varepsilon)=\Theta\left(\varepsilon_{\mathrm{b}}+\frac{\Delta}{2}-\varepsilon\right)\sum_{n=1}^{n_{0}+1}
n \delta (\varepsilon-\Delta n).\label{20a}
\end{equation}
In the intervening region, the density of states is the sum of a
piecewise continuous function and the density of a discrete
spectrum:
\begin{eqnarray}
\mathfrak{N}_{2\mathrm{pc}}(\varepsilon)&=&\Theta\left(\varepsilon-\frac{\Delta}{2}-\varepsilon_{\mathrm{b}}\right)
\Theta(2\varepsilon_{\mathrm{b}}-\varepsilon)\nonumber\\
&&{}\times
\left\{\sum_{n=n_{0}+2}^{2n_{0}+1}(2n_{0}+2-n)\delta(\varepsilon-\Delta
n)+ \frac{2L}{\pi}\sum_{i=0}^{n_{\varepsilon}}\left[\varepsilon-\Delta
\left(i+\frac12\right)\right]^{-1/2} \right\}, \nonumber\\
n_{\varepsilon}&=&\left\lceil\frac{\varepsilon}{\Delta}\right\rceil-(n_{0}+1).
\label{20b}
\end{eqnarray}
In the region of great energy values, the density of states is a
continuous function:
\begin{equation}
\mathfrak{N}_{2c}(\varepsilon)=\Theta(\varepsilon-2\varepsilon_{\mathrm{b}})\left\{\frac{2L}{\pi}\sum_{i=0}^{n_{0}}
\left[\varepsilon-\Delta\left(i+\frac12\right)\right]^{-1/2}+
\frac{2L^2}{\pi^2}\arcsin\left(1-\frac{2\varepsilon_{\mathrm{b}}}{\varepsilon}\right)\right\}.
\label{20c}
\end{equation}
\end{subequations}
These formulae are illustrated in figure~1. Let us represent
the states by the points at the first quadrant. The Cartesian
coordinates of the point are the eigenvalues of 1D Hamiltonians
that are the components of the 2D Hamiltonian. The density of
states in the point is the product of 1D densities in the
projections of the point multiplied by $\rd\varepsilon$ [see the
derivation of formula~(15)]. All the states with the same energy
$\varepsilon =\alpha$ are located at the line segment that cuts
off the intercepts  equal to $\alpha$ on the axes. The 2D density of
states that have the energy $\alpha$ is the sum of the densities
over all points of this line segment. The coordinates of the
states that are located in the square
$O\varepsilon_{1b}D\varepsilon_{2b}$ or on its sides are as follows:
$\varepsilon_{i}=\Delta(n_{i}+1/2)$. Therefore, the states that
have energy eigenvalue $\varepsilon\leqslant  \varepsilon_{\mathrm{b}}+\Delta/2$
(for example the point $A$) form the discrete spectrum [formula~
(20a)]. The degeneracy multiplicity of the discrete state with
energy $\varepsilon_{n}=n\Delta,\quad(n=1,\: 2,\ldots n_{0}+1)$ is equal
to $n$. These degenerate levels are transformed in zonule, if the
quantities $\gamma(n)$ are taken into account, but this broadening
is small everywhere except the immediate neighborhood of
$\varepsilon_{\mathrm{b}}$. Among the states with energy
$\varepsilon_{\mathrm{b}}+\Delta/2<\varepsilon<2\varepsilon_{\mathrm{b}}$ [the
section $b_{1}b_{2}$, formula~(20b)] there should be such ones that
are located outside the square
$O\varepsilon_{1b}D\varepsilon_{2b}$. One of the projections of
such a state is located in the region of the continuous 1D spectrum
(the point $B$). Therefore, the formula~(20b) consists of a
discrete part and a piecewise continuous part. The degeneracy
multiplicity of the discrete state with energy
$\varepsilon_{n}=n\Delta,\quad(n=n_{0}+2,\ldots 2n_{0}+1)$ is equal to
$2n_{0}+2-n$. When the state energy $\varepsilon>2\varepsilon_{\mathrm{b}}$
[the section $c_{1}c_{2}$, formula~(20c)] the point $C'$ is
similar to the point $B$, and for the point $C$, the both
projections are located in the regions of a continuous 1D spectrum.
\begin{figure}[!h]
\centerline{
\includegraphics[width=10cm]{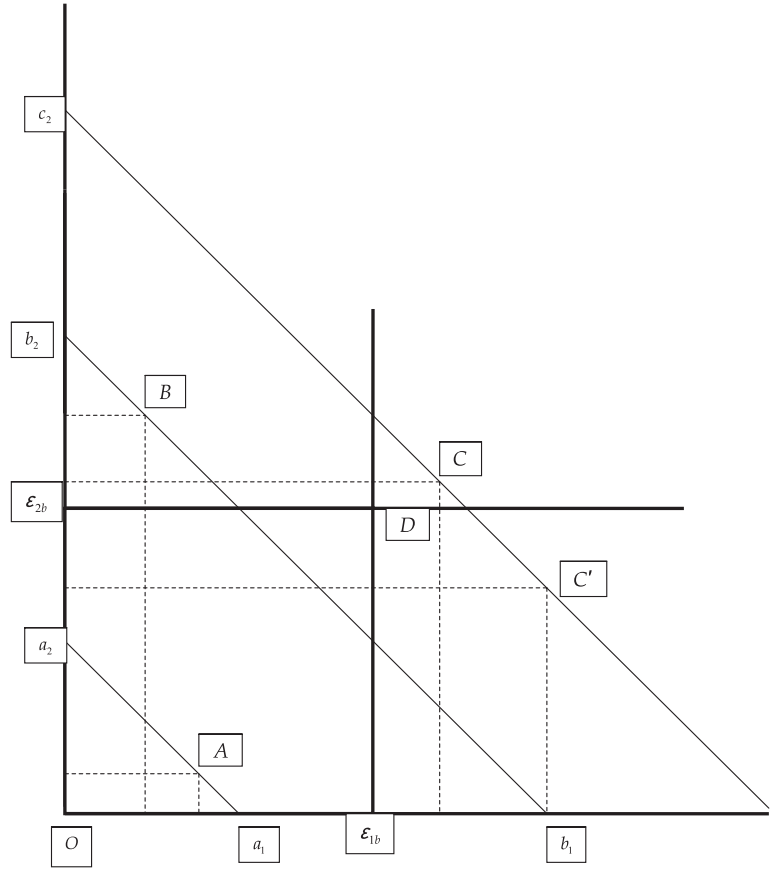}
}
\caption{The illustration of calculation of the density of states for 2D electron in the magnetic field in a square.}
\end{figure}

Let us consider the weak field case, when the Fermi energy
$\varepsilon_{\mathrm{F}}> 2\varepsilon_{\mathrm{b}}$. If the density of states is
determined by averaging over the interval $\Delta$, and the
amendments that take the  form of the area into account are neglected, the
density of states $\widetilde{\mathfrak{N}_{2}(\varepsilon)}$
should be like the one obtained in the work \cite{2}:
\begin{equation}\label{21}
\widetilde{\mathfrak{N}_{2}(\varepsilon)}=\Theta\left(\frac{m^2\omega^2
S}{4\pi\hbar^2}-\varepsilon\right)\left[\frac{\varepsilon}{\Delta^2}\right]+\Theta\left(\varepsilon-\frac{m^2\omega^2
S}{4\pi \hbar^2}\right)\left[\frac{S}{4\pi}\right].
\end{equation}
The integral of the density of states that determines the number
of states whose energies are smaller than $\varepsilon$, can
be represented as follows:
\begin{equation}\label{22}
N(\varepsilon>2\varepsilon_{\mathrm{b}})=\int_{0}^{\varepsilon}\widetilde{\mathfrak{N}_{2}(\alpha)}\rd \alpha
+\int_{0}^{\varepsilon}\left[\mathfrak{N}_{2}(\alpha)-\widetilde{\mathfrak{N}_{2}(\alpha)}\right]\rd \alpha,
\end{equation}
where $\mathfrak{N}_{2}(\alpha)$ is taken from the
formulae~(20a,~20b,~20c). The equation for calculation of the
Fermi energy $\varepsilon_{\mathrm{F}}$ is obtained by equating
$N(\varepsilon_{\mathrm{F}})$ to the total number of electrons $N_{0}$.
(For the sake of simplicity, the spin and the Pauli paramagnetism
is not considered). This equation can be considered as the
implicit definition of the function $\varepsilon_{\mathrm{F}}(H)$. If the
second term in the formula (22) is neglected, then:
\begin{equation}\label{23}
\varepsilon_{\mathrm{F}}(H)=\varepsilon_{\mathrm{F}0}+\varepsilon_{\mathrm{F}1}(H)=\varepsilon_{\mathrm{F}0}+\frac{1}{2\pi}\left(\frac{m\omega
L}{\hbar}\right)^2=\varepsilon_{\mathrm{F}0}+\frac{1}{2\pi}\left(L\Delta
\right)^2,
\end{equation}
where $\varepsilon_{\mathrm{F}0}=4\pi N_{0}/S$ is the Fermi energy in the
commonly used theory when there is no magnetic field.

The Fermi energy depends on the magnetic field due to the dependence on the magnetic induction of two parameters:
$\varepsilon_{\mathrm{F}}(H)=\varepsilon_{\mathrm{F}}\left[\Delta(H),n_{0}(H)\right]$.
These dependencies are qualitatively different:
$\Delta(\omega)=m\omega/\hbar$ is the linear function of $H$ that
can take on any values, and $n_{0}(\omega)=\left\lceil4m\omega
L^2/\pi^2\hbar\right\rceil$ can be only integer. Therefore, when
the magnetic induction varies in an interval, in which the
parameter $n_{0}$ does not vary, the singular points of the
function $\mathfrak{N}_{2}(\varepsilon)$ [formulae~(20a)--(20c)]
move continuously and
$\varepsilon_{\mathrm{F}}(H)=\varepsilon_{\mathrm{F}}\left[\Delta(H)\right]$ varies
continuously. When the  variation of the magnetic induction
changes the parameter $n_{0}$, the number of the singular points
 changes, the spectrum reconstructs, and $\varepsilon_{\mathrm{F}}$
varies non-continuously. Formula (23) that describes the function
$\varepsilon_{\mathrm{F}}(H)$ when the density of states is smoothed,
should be supplemented by the oscillatory term
$\varepsilon_{\mathrm{F}2}(\Delta ,n_{0})$. By integration
$\mathfrak{N}_{2}(\alpha)$ between the limits $0$ and
$\varepsilon_{\mathrm{F}}>2\varepsilon_{\mathrm{b}}$ we obtain:
\begin{align}
N(\varepsilon_{\mathrm{F}}>2\varepsilon_{\mathrm{b}})={}&n_{0}^2-\frac{4L\sqrt{\Delta}}{\pi}\left[n_{0}^{3/2}
-\sum_{i=0}^{n_{0}}\left(\frac{\varepsilon_{\mathrm{F}}}{\Delta}-i\right)^{1/2}\right]\notag\\
&+\frac{2L^2}{\pi^2}\left[\varepsilon_{\mathrm{F}}\arcsin\left(1-\frac{2\Delta
n_{0}}{\varepsilon_{\mathrm{F}}}\right)-2\sqrt{\Delta
n_{0}(\varepsilon_{\mathrm{F}}-\Delta n_{0})}+2\Delta n_{0}\right].\label{24}
\end{align}
If  $n_{0}(H)-n_{0}(H\pm h)=\pm 1$  then  $\Delta (H)-\Delta (H\pm
h) =\pm (\pi^2/4L^2)$. Then, the continuous change of
$\varepsilon_{\mathrm{F}}$ on this interval is:
\begin{align}
\delta_{c}\varepsilon_{\mathrm{F}} &{}=\frac{\partial
\varepsilon_{\mathrm{F}}}{\partial\Delta}\left(\pm\frac{\pi^2}{4L^2}\right)=-\frac{\partial
N}{\partial\Delta}\left(\frac{\partial
N}{\partial\varepsilon}\right)^{-1}\left(\pm\frac{\pi^2}{4L^2}\right)=
-\frac{1}{\mathfrak{N}_{2}(\varepsilon_{\mathrm{F}0})}\frac{\partial
N}{\partial\Delta}\left(\pm\frac{\pi^2}{4L^2}\right)\notag\\
&{}\approx-\frac{\pi}{L^2} \left(-\frac{n_{0}^{2}}{\Delta }
\sqrt{\frac{\varepsilon_{\mathrm{F}0}}{\Delta n_{0}}}\right)
 \left(\pm\frac{\pi^2}{4L^2}\right)
\approx\pm\frac{2}{L}\sqrt{\varepsilon_{\mathrm{F}0}}=
\pm\frac{2}{L^2}\sqrt{\pi N_{0}}\,. \label{25}
\end{align}
This change is much larger than the change that is described
by the formula (23). It describes the oscillations of
$\varepsilon_{\mathrm{F}2}(\Delta ,n_{0})$. At the end of the interval of
the continuous change of $\varepsilon_{\mathrm{F}2}(\Delta ,n_{0})$, when
$n_{0}$ changes by $\pm 1$, the jump of the function $\varepsilon_{\mathrm{F}}$
is:
\begin{equation} \label{26}
\delta_{j}\varepsilon_{\mathrm{F}}=\pm\frac{\partial
\varepsilon_{\mathrm{F}}}{\partial n_{0}}= \mp \frac{\partial N}{\partial
n_{0}} \left(\frac{\partial N}{\partial
\varepsilon_{\mathrm{F}}}\right)^{-1} =\mp\frac{\pi}{L^2}\frac{\partial
N}{\partial n_{0}}\approx \mp
\frac{\pi}{L^2}\left(\frac{n_{0}\varepsilon_{\mathrm{F}0}}{\Delta}\right)
=\mp\frac{2}{L}\sqrt{\varepsilon_{\mathrm{F}0}}\,.
\end{equation}
When the magnetic field varies, the Fermi energy oscillates with
the amplitude \;$(\hbar/L)\sqrt{2E_{\mathrm{F}0}/m}$\; and \;the\; period\;
$\delta H=\pi^2\hbar/4L^2e_{\mathrm{r}}$.\; If $L^2\backsim
10^{-4}$~m$^{2}$ \;and \;$e_{\mathrm{r}}\backsim 10^{-4}e$, this
period is of the order of $10^{-6}$~T. The monotonous change of
the Fermi energy with the magnetic field is described by the
formula (23).

It can be similarly proved that the energy of the electron gas in
the magnetic field is described by formula:
\begin{equation}\label{27}
E=E_{0}+\frac{S^2\rho_{0}e_{\mathrm{r}}^2H^2}{16\pi
m}-\frac{S^3e_{\mathrm{r}}^{4}H^4}{1536\pi^{3}\hbar^2m}+E_{\mathrm{osc}}\,.
\end{equation}
Here, $E_{0}$ is the energy of the electron gas in the absence of
the magnetic field, $\rho_{0}=N_{0}/S$ is the density of gas. This
formula differs from the energy of gas that was calculated in the
work \cite{2} by an ultimate term $E_{\mathrm{osc}}$ that describes the
oscillation of the energy with the amplitude
$(4/3\pi)\sqrt{(2mL^2/\hbar^2)E_{\mathrm{F}0}^{3}}\,$.

We take into account that the number of discrete levels $n_{0}$ is
an integer and obtain the characteristics of the gas  that have the
oscillating dependence on the magnetic field. In the commonly used
theory, the degenerate multiplicity of equidistant levels $d(H)$ is
an integer and is the same for all levels. This quantity differs from
$n_{0}$ only by the numerical coefficient that is of the order of
unity. When the magnetic field, for example, increases, the gas
energy should increase linearly until $d$ is constant. When $d$
increases by 1, the $\lceil N_{0}/d\rceil$ electrons should drop
from the top level and the energy decreases by jumps. If the number of
electrons on the top level is less than $\lceil N_{0}/d\rceil$,
the jump amplitude should be smaller. These decreased jumps should
recur, and only this oscillation is considered in the common
theory. The 2D electron gas in the magnetic field has been
considered in the monographs \cite{3,4}. The fact that
$d(H)$ is an integer is not taken into account in these works. In the
work [2] it was shown that the system of equidistant degenerated
levels cannot be a correct description of the one-particle spectrum
of the electron gas in the magnetic field because in this theory,
the angular momentum conservation and the Coulomb interaction are
not taken into account.

\section{The linear oscillator with zero boundary condition}

Let us  change the variable in the equation (17)
$z=x\sqrt{\Delta}$ and designate $u=-2E/\hbar\omega$. Then,
the equation obtains the form of a standard equation for the
function of parabolic cylinder (see handbook~\cite{5}):
\begin{equation}\label{28}
\frac{\rd^2\psi}{\rd z^2}-\left(\frac{z^2}{4}+u\right)\psi=0.
\end{equation}
The even and odd solutions of this equation are:
\begin{align}
&\psi_{\mathrm{e}}=A_{\mathrm{e}}\exp\left(-\frac{z^2}{4}\right)
\Phi\left(\frac{u}{2}+\frac{1}{4},\frac{1}{2};\frac{z^2}{2}\right);\notag\\
&\psi_{\mathrm{o}}=A_{\mathrm{o}}z\exp\left(-\frac{z^2}{4}\right)
\Phi\left(\frac{u}{2}+\frac{3}{4},\frac{3}{2};\frac{z^2}{2}\right).\label{29}
\end{align}
Here, $\Phi(a,c;t)$ is the degenerate hypergeometric function
(DHF), $A_{\mathrm{e(o)}}$ are the normalization constants. The middle of
the line segment  is zero of the coordinate. The length of the
 segment is $2L$. To satisfy the boundary condition (18), the
eigenvalue $u$ should be such that
\begin{equation}\label{30}
\Phi\left(\frac{u}{2}+\frac{1}{4},\frac{1}{2};z_{L}^2\right)=0,\quad\mbox{or}\quad
\Phi\left(\frac{u}{2}+\frac{3}{4},\frac{3}{2};z_{L}^2\right)=0;\qquad
z_{L}^2=\frac{m\omega L^2}{2\hbar}\,.
\end{equation}
This problem could not be consequently considered because in the
description of nulls of DHF in all mathematical handbooks (see for
example~\cite{5},\cite{6}) an inaccuracy takes place. It is
proved that, if $a<0$ and $c>0$, the number of nulls of the
function $\Phi(a,c;t)$ is equal to $(-a)$, if it is integer, and
is equal to $\lceil -a\rceil +1$, if $(-a)$ is non-integer. It is
 also proved that nulls are described approximately by the formula:
\begin{equation}\label{31}
\xi_{i}(a,c)=\frac{1}{2c-4a}j_{c-1,i}^2 \,,
\end{equation}
if $|a|\gg 1$. Here, $\xi_{i}(a,c)$ is the null of DHF that has the
number $i\leqslant \lceil -a\rceil +1$ in the order of increasing,
$j_{c-1,i}^2$ is the square of the respective null of the Bessel
function of the first kind $J_{c-1}(x)$. If this were so, the
greatest nulls of the functions that are the solutions of the
considered problem should be linear functions of the eigenvalue $u$.
Then, the smallest eigenvalues should be $\sim m\omega L^2/2\hbar$,
i.e., the spectrum should begin from very large values of energy.
In the work \cite{7} (see also \cite{2}) it was shown that the
formula (31) describes each null of the DHF only when $(-a)$ is
integer. Then, the Kummer power series that describes DHF
terminates at the term with number $1-a$, and the DHF in the
formulae (30) are proportional to the Laguerre polynomials
$L_{(-a)}^{\pm 1/2}(z_{L}^2)$. The number of nulls of these
polynomials is $k=-a$. All these nulls are real, positive, simple
and are described by the formula~(31). Let us consider the DHF
$\Phi(a,1/2;t)$ when $a=-k-\gamma$, where $k$ is an integer and
$0<\gamma<1$. Then, this DHF has $k+1$ nulls of which $k$ nulls
come out of the nulls of Laguerre polynomial that are diminished
by quantities which are proportional to $\gamma$. These nulls are
described by the formula (31). The null that has the number $k+1$
is the largest, and its dependence on $\gamma$ should possess the following properties:
\begin{equation} \label{32}
\lim_{\gamma \to 0}\left[\xi_{k+1}(-k-\gamma,1/2)\right]\to\infty
,\qquad \lim_{\gamma \to
1}\left[\xi_{k+1}(-k-\gamma,1/2)\right]=\xi_{k+1}(-k-1,1/2).
\end{equation}
To calculate $\xi_{k+1}(-k-\gamma,1/2)$ when $\gamma$ is small,
the DHF $\Phi(-k-\gamma ,1/2;t)$ should be changed by the
asymptotic expression. The Kummer power series would be
represented as follows:
\begin{equation}\label{33}
\Phi(-k-\gamma ,1/2;t)=P_{k}^{1/2}(\gamma ,t)-\gamma
Q_{k}^{1/2}(\gamma ,t)-\gamma(1-\gamma)T_{k}^{1/2}(\gamma ,t).
\end{equation}
Here, $P_{k}^{1/2}(\gamma ,t)$ is a polynomial that can be obtained
from the Laguerre polynomial $L_{k}^{-1/2}(t)$ by changing $k\to
k+\gamma$ in its coefficients, $\gamma Q_{k}^{1/2}(\gamma ,t)$ is
the next term of the Kummer series that is proportional to
$\gamma$, $\gamma (1-\gamma )T^{1/2}_{k}(\gamma , t)$   is the
remaining infinite series that also is proportional to $\gamma$.
It can be shown that when $t$ is large and $\gamma$ is small
\begin{equation}\label{34}
\gamma (1-\gamma)T_{k}^{1/2}(\gamma ,t)\approx \gamma
(-1)^{k}(1+\gamma)_{k}\sqrt{\pi}(t)^{-n-1/2}\exp(t).
\end{equation}
Here, $(b)_i=b(b+1)\cdots(b+i-1)$ is the Pochhammer symbol.
The second term in the formula (33) can be neglected. The
polynomial $P_{k}^{1/2}(\gamma ,t)$ can be changed by the last
term, when $t$ is large:
\begin{equation}\label{35}
P_{k}^{1/2}(\gamma ,t)\approx
\frac{(-k-\gamma)_{k}t^{k}}{(1/2)_{k}k!}=(-1)^{k}\frac{(1+\gamma)_{k}t^{k}}{(1/2)_{k}k!}\,.
\end{equation}
We set the obtained approximate expression for DHF equal to zero.
This equation defines the quantity $\gamma$ as the function of
$\xi_{k+1}(-k-\gamma ,1/2)=t_{0}$:
\begin{equation}\label{36}
\gamma_{1/2}(k,t_{0})=\left[k!\Gamma
\left(k+\frac12\right)\right]^{-1}t_{0}^{2k+1/2}\exp(-t_{0}).
\end{equation}
For the boundary condition to be satisfied by the
largest null of DHF $\Phi(-k-\gamma,1/2;t)$, this null should be
larger than $\xi_{k+1}(-k-1,1/2)$, i.e.,
\begin{equation}\label{37}
t_{0}=\frac{m\omega L^2}{2\hbar}>\frac{1}{1+4(k+1)}j_{-1/2,k+1}^2\,,
\end{equation}
$j_{-1/2,l}=\pi(l-1/2)$, where $l$ is integer. Therefore, the
boundary condition can be satisfied by the largest null of DHF, if
the value of $k$ is not greater than $k_{0}$:
\begin{equation}\label{38}
k_{0}\approx\left\lceil\frac{2m\omega L^2}{\pi^2\hbar}\right\rceil
.
\end{equation}
This is the approximate formula, but the operation of taking
an integer part emphasizes that $k_{0}$ is an integer, and when the
frequency changes, this quantity does not change continuously and
takes only integer values. The eigenvalues for the even solutions:
\begin{equation}\label{39}
E_{\mathrm{e}}=\hbar\omega\left(k_{\mathrm{e}}+\frac{1}{4}+\gamma_{1/2}(k_{\mathrm{e}})\right).
\end{equation}
If $k>k_{0}$, the boundary condition can be satisfied by one of the
nulls of DHF that is described by the formula (31) when
$(-a)>k_{0}$. To calculate these values $a$, i.e., the
eigenvalues of energy, we use the first term of the expansion DHF
over the Bessel functions (see monograph \cite{6}). This expansion
is rapidly convergent, when $|a|$ is large. We obtain
the following even solution of the equation (28):
\begin{align}\label{40}
\psi_{\mathrm{e}}\approx
A_{\mathrm{e}}\Gamma\left(\frac{1}{2}\right)\left(-\frac{uz^2}{4}\right)^{1/4}\!\!
J_{-1/2}\left(2\sqrt{-uz^2/4}\right)=A_{\mathrm{e}}\cos\left(z\sqrt{2E_{\mathrm{e}}/\hbar\omega}\right)=
A_{\mathrm{e}}\cos\left(\frac{x}{\hbar}\sqrt{2mE_{\mathrm{e}}}\right).
\end{align}
The eigenvalues of the energy for which these wave functions are equal to
zero at the ends of a segment are obtained:
\begin{equation}\label{41}
E_{\mathrm{e}}=\frac{\pi^2\hbar^2}{2mL^2}\left(k_{\mathrm{e}}+\frac{1}{2}\right)^2=\frac{\pi^2\hbar^2}{8mL^2}(2k_{\mathrm{e}}+1)^2,\qquad
k_{\mathrm{e}}>k_{0}\,.
\end{equation}
A similar computation for the odd wave function leads to the
results:
\begin{align}\label{42}
&\gamma_{3/2}(k,t_{0})=\left[k!\Gamma\left(k+\frac{3}{2}\right)\right]^{-1}t_{0}^{2k+3/2}\exp(-t_{0}),\qquad
E_{\mathrm{o}}=\hbar\omega\left(k_{\mathrm{o}}+\frac{3}{4}+\gamma_{3/2}(k_{\mathrm{o}})\right);\\
&\psi_{\mathrm{o}}\approx
A_{\mathrm{o}}\sin\left(\frac{x}{\hbar}\sqrt{2mE_{\mathrm{o}}}\right),\qquad
E_{\mathrm{o}}=\frac{\pi^2\hbar^2}{2mL^2}k_{\mathrm{o}}^2=\frac{\pi^2\hbar^2}{8mL^2}(2k_{\mathrm{o}})^2,\quad
k_{\mathrm{o}}>k_{0}\,.\notag
\end{align}
The formulae for the spectrum can be unified, if it is taken into
account that the amendments $\gamma$ can be neglected everywhere
except the immediate neighborhood of $k_{0}$:
\begin{equation}\label{43}
E_{n}=\frac{\hbar\omega}{2}\left(n+\frac{1}{2}\right),\qquad
n<n_{0}=2k_{0};\qquad E_{n}=\frac{\pi^2\hbar^2}{8mL^2}n^2,\qquad
n>n_{0}\,.
\end{equation}
If in the equation (17) we change $\omega=2\omega_{0}$, this
equation will gain the form of a standard equation for the
linear oscillator with frequency $\omega_{0}$. If $L\to\infty$,
the boundary condition (18) should be changed by the requirement of
normability of the wave functions. Then, the obtained solution of
the problem turns into the common solution for the linear
oscillator. When $n>n_{0}$, the approximate solutions (40) and (42)
are the common solutions for the particle in the rectangular
potential well. However, the energy $E_{\mathrm{b}}=\hbar\omega
n_{0}/2=2m\omega^2L^2/\pi^2$ that is the boundary between the two
kinds of solutions does not coincide with the value of the
potential energy at the boundary of the area as might be expected from
the quasiclassical consideration.

\ukrainianpart

\title{Густина одночастинкових станів для 2D електронного газу у магнітному полі}
\author{І.М.~Дубровський}
\address{Інститут металофізики, бульв. Вернадського 36, Київ 03680, Україна}

\makeukrtitle

\begin{abstract}
\tolerance=3000%
Густина станів частинки у 2D області не залежить від енергії і форми області тільки при великих значеннях енергії. При малій енергії густина станів у прямокутній потенціальній ямі суттєво залежить від форми області. Якщо дно потенціальної ями має потенціальний рельєф, то він може визначати малі влас\-ні значення енергії як дискретні рівні. У цьому випадку розміри і форма області не мають значення. Якщо приймати до уваги збереження нульового значення кутового моменту, ефективний одночастинковий Гамільтоніан для 2D електронного газу у магнітному полі у колі є Гамільтоніаном з параболічним потенціалом і відбиваючими границями. Припускається, що у квадраті Гамільтоніан має такий самий вигляд. 2D густина станів у квадраті може бути обчислена як згортка 1D густин. Обчислено густину станів 2D електронного газу у магнітному полі. Вона складається з трьох областей. Коли енергії малі, спектр є дискретним. У проміжній області густина станів є сумою проміжково-неперервної функції і густини дис\-крет\-но\-го спектру. При великих значеннях енергії густина станів є неперервною функцією енергії. Одержано залежність енергії Фермі від магнітного поля, коли поле є слабким і енергія Фермі знаходиться в області неперервного спектру. Енергія Фермі має доданок, який осцилює і, в серед+ньому, зростає пропорційно квадрату магнітної індукції. Повна енергія електронного газу у магнітному полі також осцилює  і зростає, коли магнітне поле монотонно збільшується.
\keywords густина станів, електронний газ, магнітне поле, енергетичний спектр, енергія Фермі, повна енергія

\end{abstract}

\end{document}